\documentclass[sigconf]{acmart}

\usepackage[T1]{fontenc}
\usepackage[utf8]{inputenc}
\usepackage{amsmath}

\usepackage{amssymb}
\usepackage{booktabs}

\usepackage{graphicx}
\usepackage{fontawesome}
\usepackage{listings}
\usepackage{multirow}
\usepackage{stfloats}
\usepackage{tabularx}
\usepackage{url}
\usepackage{xcolor}
\usepackage{tikz}
\usetikzlibrary{arrows.meta,positioning,fit,matrix,calc,shapes.geometric}

\newcommand{\claude}{Claude Opus 4.7}
\newcommand{\codex}{Codex GPT-5.5}
\newcommand{\lcb}{LiveCodeBench}
\newcommand{\sid}{\ensuremath{\mathrm{SID}}}
\colorlet{condA}{blue!70!black}
\colorlet{condO}{red!70!black}
\colorlet{condAO}{magenta!70!black}
\colorlet{condOA}{green!45!black}
\colorlet{condAA}{cyan!45!black}
\colorlet{condOO}{black!65}

\newcommand{\promptkey}[1]{\textcolor{blue!65!black}{\textbf{#1}}}
\newcommand{\promptplaceholder}[1]{\textcolor{cyan!40!black}{\texttt{#1}}}
\newcommand{\prompttag}[1]{\textcolor{magenta!65!black}{\texttt{\textbf{#1}}}}
\lstdefinestyle{prompttemplate}{
    basicstyle=\ttfamily\footnotesize,
    backgroundcolor=\color{black!2},
    breakatwhitespace=true,
    breaklines=true,
    belowcaptionskip=0.8em,
    columns=fullflexible,
    frame=single,
    keepspaces=true,
    rulecolor=\color{black!25},
    escapeinside={(*@}{@*)}
}
\usepackage{caption}

\title{Cross-Model LLM Code Review: Should you use Claude to review Codex or vice versa?}

\author{Zuodong Xiang}
\affiliation{%
  \institution{University of California, Davis}
  \country{USA}}
\email{shawnzxiang@gmail.com}

\author{Yike Zhang}
\affiliation{%
  \institution{Johns Hopkins University}
  \country{USA}}
\email{zykk0330@gmail.com}

\author{YueMing Zhang}
\affiliation{%
  \institution{California State University, Long Beach}
  \country{USA}}
\email{Simon.Zhang01@student.csulb.edu}

\author{Hailu Xu}
\affiliation{%
  \institution{California State University, Long Beach}
  \country{USA}}
\email{hailu.xu@csulb.edu}

\thanks{This work is partially supported by US NSF Award \#2436427.}

\acmConference[Agentic SE @ KDD '26]{Agentic Software Engineering (SE 3.0): The Rise of AI Teammates}{August 10, 2026}{Jeju, Korea}
\acmYear{2026}
\copyrightyear{2026}

\begin{document}
\begin{abstract}
Developers increasingly use two coding agents together: one writes a draft, and the other reviews it. However, it is not clear whether the pairing is worth its cost and time, or whether the order of the pairing matters. We run a controlled experiment on 116 recent hard and medium \lcb{} tasks with \claude{} and \codex{} across six conditions to approximate a software practitioner's workflow: both solo baselines, both cross-model orderings, and both same-model orderings. The reviewer sees the problem and the writer's draft but cannot execute tests, which approximates a code review step. \claude{} review raises \codex{} drafts from 71.6\% to 89.7\% ($p_{BH}=.001$); \codex{} self review raises them to 84.5\% ($p_{BH}=.022$). The reverse direction does not pay off: \codex{} reviewing \claude{} drafts drops the pass rate from 91.4\% to 82.8\% ($p_{BH}=.046$), and \claude{} self review leaves the 91.4\% baseline unchanged. Our evaluation indicates that the useful pairing is asymmetric: use \claude{} to review \codex{}, not the other way around.
\end{abstract}



\maketitle

\section{Introduction}
\label{sec:intro}
Many coding workflows now involve more than a single LLM call. A developer asks one agent to write a draft and a second agent to review it before the change lands, and the second pass starts to behave like a software engineer reviewing code. The two arrangements we see most often in practice are {\claude}~\cite{anthropic2025claudeopus4} writing with {\codex}~\cite{openai2025gpt5} reviewing, and the reverse, with {\codex} writing and {\claude} reviewing. Each pairing can change the workflow's final result. A different model, commonly equipped with different knowledge and expertise, is now responsible for catching things that the first one missed. But it can also cause regressions, replace a working solution with a worse one, or simply add cost and latency without measurable benefit.

Today's code generation leaderboards rank single models in isolation. They do not address which of two available agents should write, which should review, whether the second pass is worth its cost, or which direction the pairing should run to yield the best correctness. \claude{} writing and \codex{} reviewing is not the same workflow as the reverse, even though the two models involved are identical.

We study this as a role assignment problem between a writer and a reviewer. The setup approximates a software engineer's workflow: a single review pass with no execution feedback, so the measured effect is the review intervention itself rather than tool use, iteration, or memory of earlier attempts. The reviewer reads the problem statement and the writer's draft and writes a final program; it cannot run the code or query a test runner. This is also commonly referred to as \emph{static review}: the pre-CI inspection step a developer or code review bot performs before any tests run.

The experiments use \lcb{}, a continuously updated benchmark of self contained competitive programming problems sourced from recent online judges, designed to reduce the risk of contamination in code evaluation \cite{jain2025livecodebench}. All problems we use were released after 2025, after the training cutoffs of \claude{} and \codex{}, so neither model could have memorized their solutions during training. Each task is a single file Python problem with a problem statement, optional starter code, and a hidden test suite. Tasks carry difficulty labels (easy, medium, hard), and we use only the hard and medium difficulty tiers to approximate real life software engineering tasks worth using agents. We pool the two difficulty tiers and compare six conditions: \claude{} solo, \codex{} solo, Claude output reviewed by Codex, Codex output reviewed by Claude, Claude output reviewed by Claude, and Codex output reviewed by Codex. \lcb{} fits this question well because the same problem can be evaluated repeatedly under different writer and reviewer assignments against a common hidden test evaluator, which isolates the review pass without changing the task distribution. All writer and reviewer turns use the vendors' \emph{high} reasoning effort setting, which uses more tokens, cost, and latency than medium effort. We report pass rate together with reviewer fixes, regressions, latency, and cost.

In this paper, we focus on the following questions:

\begin{list}{}{\leftmargin=0pt \labelwidth=0pt \labelsep=0pt \itemindent=0pt \itemsep=0.35em \parsep=0pt \topsep=0.35em}
\item \textbf{RQ1.} Does a reviewer pass improve overall correctness?
\item \textbf{RQ2.} Does the value of a reviewer pass depend on which model writes and which model reviews?
\item \textbf{RQ3.} When does the reviewer step fix the draft, and when does it regress it?
\item \textbf{RQ4.} Are correctness gains worth added cost and latency?
\end{list}

Review effectiveness in our experiments turns out to be asymmetric. \claude{} is the strongest solo writer in the sample, \claude{} reviewing \codex{} drafts recovers most of the gap between \codex{} solo and \claude{} solo, and \codex{} reviewing \claude{} drafts makes the pipeline worse. We analyze this asymmetry through paired correctness tests, fix/regression counts, and cost-latency trade-offs, then derive a workflow recommendation: review is most useful when the initial draft comes from \codex{}.

\section{Background and Related Work}

\subsection{Code generation benchmarks}
Code generation evaluation has usually been organized around standalone model performance. HumanEval made functional correctness a central measure for generated code \cite{chen2021evaluating}, and MBPP extended this style of evaluation to short Python programming tasks \cite{austin2021program}. APPS moved the setting toward harder programming-challenge problems \cite{hendrycks2021measuring}, and AlphaCode showed that large-scale generation and filtering can reach nontrivial competitive programming performance \cite{li2022competition}. \lcb{} addresses benchmark saturation and contamination by collecting recent coding contest problems over time from LeetCode, AtCoder, and Codeforces, tagging each problem with its public release date \cite{jain2025livecodebench}. The release date metadata lets evaluators score a model only on problems published after its training cutoff, which controls for the data leakage documented on older static benchmarks. \lcb{} also reports several evaluation scenarios beyond code generation, including self repair, code execution, and test output prediction. 

Prior work also argues that quality should be reported together with efficiency rather than in isolation \cite{zhang2026preferenceefficiency,jiang2026drpdistilledreasoningpruning,gao2026dspc,shen2026efficiency}. This matters even more for review workflows because a reviewer pass adds another model call, more tokens, and more latency. We therefore treat pass rate, cost, latency, fixes, and regressions as a set of measurements rather than a leaderboard number.

\subsection{Iterative refinement and self debug}
Self-Refine and Reflexion study a single model that improves its own output through feedback turns \cite{madaan2023self,shinn2023reflexion}. Self-Debug teaches a model to repair its own code from execution traces \cite{chen2024teaching}. Olausson et al.\ ask whether self-repair is a silver bullet for code generation and find that the gain depends on whether the repair model is meaningfully stronger than the draft model \cite{olausson2024self}. On HumanEval and APPS with Code Llama, GPT-3.5, and GPT-4, same model self repair produces only modest accuracy improvements because it is bottlenecked by the model's ability to provide useful feedback on its own code; the largest gains appear when a stronger model repairs a weaker model's drafts, and human level debugging substantially outperforms self generated feedback. Our same model conditions will reproduce this finding but with execution feedback removed. Recent work also shows that memory can change LLM agent behavior \cite{liu2026memory}, and that LLM outputs exhibit non-trivial variance across repeated runs even under fixed conditions \cite{zhou2026accuracystabilityrepeatedrunreliability, guo2025lightllm}. Our experiment isolates a related but distinct case: the reviewer is a different model, runs once, has no execution feedback, and has no memory of prior attempts. This removes iteration and tool use, so that the measured effect can be attributed to the cross-model review pass itself.

\subsection{Multi agent and agentic coding}
CodeT and LEVER use generated tests or learned verification to improve code outputs \cite{chen2022codet,ni2023lever}. Multi agent debate work shows that two language-model instances can improve factuality or divergent thinking by critiquing each other \cite{du2024improving,liang2024encouraging}; this is a natural framing of two models reviewing one another, although that line studies open-ended reasoning rather than executable programs. Multi-LLM combinations have also been used in domain forecasting tasks such as financial sentiment \cite{zhang2026finsentllm}, robot tasks \cite{zhang2025survey, zhang2026swiftbot}, and LLMs have been integrated with reinforcement learning for portfolio optimization \cite{wang2025gemsllm}. ChatDev and MetaGPT assign software-development roles across multiple agents \cite{qian2024chatdev,hong2024metagpt}. AgentCoder and CodeCoR study role-based coding workflows \cite{huang2023agentcoder,pan2025codecor}. Other work studies capability composition across multi-agent pipelines from a safety angle \cite{jiang2026chaincapscompositionsafetoolusingagents}. In contrast, our workflow has one writer, one static reviewer, no generated tests, and no execution feedback for the reviewer.

\subsection{Code review and LLM reviewers}
Modern code review supports defect detection, knowledge transfer, and shared code quality \cite{sadowski2018modern,bacchelli2013expectations}. In human teams, review is not only a bug finding and for readability, it is also a coordination point where a second engineer decides whether to ship as is, lightly edit, or reject a proposed change. This role structure is similar to what multi agent coding workflows are beginning to imitate when one LLM drafts code and another LLM inspects it.

Prior work studies review-comment generation and code refinement \cite{tufano2021towards,li2022automating}. Recent LLM-focused work finds that apparent success on code refinement often reflects memorization rather than comprehension \cite{lin2025codereviewqa}, and that LLM reviewers can systematically overcorrect on requirement conformance \cite{jin2026llms}. These findings motivate a stricter evaluation target for LLM review. It is not whether the reviewer produces plausible comments, but whether the final submitted program is more correct than the writer's original draft. We measure the correctness of the final program after a review pass and separately count reviewer fixes and regressions.

\section{Methodology}
\subsection{Task Setting and Prompts}
Each task is a hard or medium \lcb{} code generation problem. We did not include the easy category given they are trivial for LLMs and software practitioners to solve at this point. The writer receives the problem statement, any starter code, and is asked for a complete Python solution inside \texttt{<solution>} tags.

\begin{lstlisting}[float=t, caption={Writer prompt template},label={lst:writer-prompt}]
You are an expert competitive programmer. Solve the following programming problem.

(*@\promptkey{PROBLEM:}@*)
(*@\promptplaceholder{\{problem\}}@*)
(*@\promptplaceholder{\{starter\_code\}}@*)

Your task:
1. Produce a complete, runnable Python solution
2. Handle edge cases; optimize first for correctness, then for performance
3. Output only the final solution inside (*@\prompttag{\textless solution\textgreater}@*) tags

(*@\prompttag{\textless solution\textgreater}@*)
# final python code here
(*@\prompttag{\textless/solution\textgreater}@*)
\end{lstlisting}

\begin{lstlisting}[float=t, caption={Reviewer prompt template},label={lst:review-prompt}]
You are an expert code reviewer. Review the following solution to a competitive programming problem.

(*@\promptkey{PROBLEM:}@*)
(*@\promptplaceholder{\{problem\}}@*)
(*@\promptplaceholder{\{starter\_code\}}@*)

(*@\promptkey{SUBMITTED SOLUTION:}@*)
(*@\promptplaceholder{\{writer\_output\}}@*)

Your task:
1. Identify any bugs, incorrect logic, missing edge cases, or inefficiencies
2. Produce a final corrected solution - either the original if it is correct, or an improved version
3. Do NOT run or test the code; reason purely from code inspection

Output only the final solution inside (*@\prompttag{\textless solution\textgreater}@*) tags. Do not include explanations outside the tags.
(*@\prompttag{\textless solution\textgreater}@*)
# final python code here
(*@\prompttag{\textless/solution\textgreater}@*)
\end{lstlisting}

In a solo baseline, a parsed writer draft is submitted directly to the \lcb{} evaluator. In a reviewed condition, the same writer artifact becomes the submitted solution shown to a reviewer model, which also receives the original problem and starter code. The reviewer is instructed to inspect the draft for bugs, missing edge cases, incorrect logic, and inefficiencies, then write a final program inside \texttt{<solution>} tags: either the original draft if it is correct or a corrected version if it is not. The reviewer cannot run the code, query a test runner, see hidden tests, or inspect execution traces. This is the \emph{static review} setting introduced in Section~\ref{sec:intro}. All final artifacts are parsed through the same tag-based extractor before evaluation, so the measured difference between conditions is the added review pass rather than a change in evaluator or output parser. The prompt templates are fixed across models and conditions. The artifact includes the exact prompt files and per task prompt hashes.

Listings~\ref{lst:writer-prompt} and Listing~\ref{lst:review-prompt} define the two role-specific prompt templates used throughout the study. The writer prompt instructs the model to solve the competitive programming task from the provided problem statement and starter code, prioritizing correctness, edge-case handling, and computational efficiency. To ensure a uniform output artifact, the writer is required to return only a complete runnable Python solution enclosed within \texttt{<solution>} tags.

The reviewer prompt receives the same task specification together with the writer generated solution. Rather than providing feedback or commentary, the reviewer performs a static code review and must return a final executable program. Reviewers are explicitly instructed to reason solely from code inspection and are not allowed to execute or test the program.

Both prompts enforce an identical output format consisting exclusively of a Python solution wrapped in \texttt{<solution>} tags. This design removes variation arising from the explanatory text and allows all experimental conditions to be evaluated using the same downstream extraction and scoring pipeline. Task-specific content is injected through placeholders corresponding to the problem statement, the starter code, and, when applicable, the draft produced by the writer. Holding the task specification and output format constant while varying only the role assignment and model identity isolates the effects of review and cross-model collaboration.

\begin{table}[!t]
\centering
\caption{Experimental conditions. All turns use high reasoning effort.}
\label{tab:conditions}

\resizebox{\columnwidth}{!}{%
\begin{tabular}{llll}
\toprule
ID & Writer & Reviewer & Role type \\
\midrule
A & \claude{} & -- & Solo baseline \\
O & \codex{} & -- & Solo baseline \\
AO & \claude{} & \codex{} & Cross-model \\
OA & \codex{} & \claude{} & Cross-model \\
AA & \claude{} & \claude{} & Same-model \\
OO & \codex{} & \codex{} & Same-model \\
\bottomrule
\end{tabular}
}
\end{table}

\subsection{Conditions}
Table~\ref{tab:conditions} lists the six conditions. Throughout the paper we use compact labels: "A" for Anthropic \claude{}, "O" for OpenAI \codex{}. 

Two-letter labels are in writer then reviewer order, so AO is Claude output reviewed by Codex, and OA is Codex output reviewed by Claude. The solo baselines give us standalone performance. The two same-model conditions tell us whether a second pass helps at all when there is no model diversity. The two cross-model conditions answer the model role assignment question and let us see whether direction matters.

\subsection{Execution and Cost Accounting}
The runs used \texttt{claude-opus-4-7} through Claude Code 2.1.50 and \texttt{gpt-5.5} through Codex CLI. Both runs set \texttt{reasoning.effort = high}. Each turn had a 3{,}000-second CLI timeout. A timed-out turn was retried once with a doubled timeout. A task that still ended incomplete (CLI error, timeout, parse failure, or empty final artifact) was retried up to four attempts. The harness records raw outputs, final artifacts, prompt hashes (SHA-256), CLI versions, timestamps, parse metadata, retry metadata, token counts, costs, and evaluation results. API costs are computed from token counts and a pricing configuration pulled from the official websites. 

\subsection{Sample, Metrics, and Statistical Tests}
\label{sec:metrics}
We combine the complete case portions of two \lcb{} runs: one hard slice and one medium slice. \emph{Complete case analysis} restricts the data to tasks for which all six conditions produced a valid evaluated artifact: a non-null evaluation result with no CLI error, no timeout, no parse failure, no budget-exceeded flag, and a non-empty final program. The pooled complete case analysis contains 116 tasks, with task identifiers namespaced by difficulty before pooling. This is a bounded controlled sample, not a leaderboard-scale evaluation; the artifact retains the per task metadata for future audits.

\begin{table*}[t]
\centering
\caption{Pooled complete case metrics across hard and medium \lcb{} tasks. \sid{} is the change in Pass Rate relative to the writer's solo baseline; Reg.\ rate is the share of tasks where the writer-only baseline passed but the reviewed condition failed.}
\label{tab:completecase}
\begin{tabular}{lrrrrrr}
\toprule
Condition & Pass & Pass Rate [95\% CI] & \sid{} & Cost/task & Latency & Reg.\ rate \\
\midrule
\multicolumn{7}{l}{\emph{Baselines}}\\
A: \claude{} solo & 106 & .914 [.862,.957] & -- & \$.226 & 86.2s & -- \\
O: \codex{} solo & 83 & .716 [.629,.793] & -- & \$.190 & 38.5s & -- \\
\midrule
\multicolumn{7}{l}{\emph{Reviewed conditions}}\\
AO: \claude{} reviewed by \codex{} & 96 & .828 [.759,.897] & $-.086$ & \$.382 & 118.0s & .112 \\
OA: \codex{} reviewed by \claude{} & 104 & .897 [.836,.948] & $+.181$ & \$.443 & 112.4s & .043 \\
AA: \claude{} reviewed by \claude{} & 106 & .914 [.862,.957] & $+.000$ & \$.389 & 135.8s & .026 \\
OO: \codex{} reviewed by \codex{} & 98 & .845 [.776,.905] & $+.129$ & \$.312 & 67.9s & .052 \\
\bottomrule
\end{tabular}
\end{table*}

Let $y_{tc}\in\{0,1\}$ indicate whether task $t$ passed under condition $c$. The pass rate is
\begin{equation}
PR_c = \frac{1}{|T_c|}\sum_{t\in T_c} y_{tc}.
\end{equation}
For a two-pass condition $c$ with writer model $w$, the improvement delta compares the condition to the corresponding solo writer baseline $b(w)$:
\begin{equation}
\sid_c = PR_c - PR_{b(w)}.
\end{equation}
We also report average cost per task, latency, and reviewer-induced regressions. Regressions are tasks where the writer-only baseline passes but the reviewed condition fails. Reviewer fixes are tasks where the writer-only baseline fails but the reviewed condition passes.

We report bootstrap 95\% confidence intervals for complete case pass rate. The McNemar test~\cite{mcnemar1947note} compares two binary outcomes on the same set of items by counting only the items where the two conditions disagree; it is the standard paired test for matched 0/1 outcomes such as task-level pass/fail across two conditions. For conditions $a$ and $b$, let
\begin{equation}
n_{10}=\sum_t \mathbf{1}\{y_{ta}=1,y_{tb}=0\}, \quad
n_{01}=\sum_t \mathbf{1}\{y_{ta}=0,y_{tb}=1\}.
\end{equation}
Under the null that the two conditions are equally likely to win on discordant tasks, $X\sim\mathrm{Binom}(n_{10}+n_{01},1/2)$. We use the exact two-sided McNemar p-value,
\begin{equation}
\begin{aligned}
p=\min\Bigl(1,\;2\min\bigl[&
P\{X\leq \min(n_{10},n_{01})\},\\
&P\{X\geq \max(n_{10},n_{01})\}
\bigr]\Bigr).
\end{aligned}
\end{equation}
For paired correctness comparisons, we apply BH (Benjamini-Hochberg) correction~\cite{benjamini1995controlling} across all pairwise contrasts generated by the analysis script. If $p_{(1)}\leq\cdots\leq p_{(m)}$ are the sorted p-values, the adjusted value is
\begin{equation}
q_{(i)}=\min_{j\geq i}\left(1,\frac{m}{j}p_{(j)}\right),
\end{equation}
with monotone adjustment from largest to smallest p-value. Table~\ref{tab:mcnemar} reports the contrasts most relevant to the role assignment argument. To understand whether the apparent writer and reviewer interaction is statistically supported beyond the per writer effects, the artifact also reports a direct AO versus OA paired contrast and the underlying per task pass vectors.

\section{Results}
Model inference ran on vendor hosted infrastructure through Claude Code and Codex CLI. The program is ran locally using a MacBook Air with Apple M4 chip and 16 GB RAMP (2025) that handled orchestration, file I/O, logging, parsing, and wall-clock timing of the remote calls.

The results section is organized around the four research questions. Table~\ref{tab:completecase} gives the complete assessment for every condition; Table~\ref{tab:heatmap} collapses those rows into the writer-by-reviewer view; Table~\ref{tab:mcnemar} tests paired differences on the same 116 tasks; and Figures~\ref{fig:outcomes}--\ref{fig:tradeoff} separate correctness gains from the cost of obtaining them. Table~\ref{tab:caseexamples} shows various inspected cases that contained fixes and regressions.

\subsection{RQ1: Does a reviewer pass improve correctness?}
Table~\ref{tab:heatmap} summarizes the headline result and Table~\ref{tab:completecase} gives the underlying numbers. The main takeaway is that the answer to RQ1 changes depending on who wrote the draft and reviewed the code. \codex{} solo lands at 71.6\%; the same drafts run through \claude{} review jump to 89.7\%, and through \codex{} review to 84.5\%. \claude{} solo is already at 91.4\%; \claude{} reviewing itself stays at 91.4\%, and handing \claude{}'s draft to \codex{} actually drops it to 82.8\%.

The size of the two positive review effects also shows that the reviewer's contribution is bounded by its relative capability rather than guaranteed by simply adding a second pass. OA adds 21 net passes over \codex{} solo (104 versus 83 passing tasks), putting a workflow that begins with a \codex{} draft within two passing tasks of \claude{} solo. OO adds 15 net passes over \codex{} solo while keeping cost and latency lower than OA. In contrast, AO loses 10 net passes relative to \claude{} solo and has the highest regression rate in Table~\ref{tab:completecase}.

\begin{table}[t]
\centering
\caption{Pass rate for each writer and reviewer pair, with the delta against the writer's solo baseline. Rows: who wrote. Columns: who reviewed. Claude as reviewer is always helpful or neutral; Codex as reviewer is helpful only when the draft is its own. }
\label{tab:heatmap}
\begin{tabular}{lcc}
\toprule
                      & \multicolumn{2}{c}{Reviewer} \\
\cmidrule(lr){2-3}
Writer                & Claude               & Codex \\
\midrule
Claude (solo 91.4\%)  & 91.4\% ($\pm 0$\,pp) & 82.8\% ($-8.6$\,pp) \\
Codex  (solo 71.6\%)  & 89.7\% ($+18.1$\,pp) & 84.5\% ($+12.9$\,pp) \\
\bottomrule
\end{tabular}
\end{table}

\begin{table}[t]
\centering
\caption{Selected complete case McNemar tests over the shared 116 tasks. Each row compares two conditions on the same set of tasks. ``First only'' is the number of tasks the first condition passed and the second failed; ``Second only'' is the reverse. $p_{BH}$ is the BH adjusted p-value across all pairwise contrasts.}
\label{tab:mcnemar}
\begin{tabular}{lrrrr}
\toprule
Comparison & First only & Second only & $p$ & $p_{BH}$ \\
\midrule
\multicolumn{5}{l}{\emph{Per writer reviewer effects}}\\
OA vs O & 26 & 5 & .0002 & .0010 \\
OO vs O & 21 & 6 & .0059 & .0222 \\
AO vs A & 3 & 13 & .0213 & .0456 \\
AA vs A & 3 & 3 & 1.0000 & 1.0000 \\
AA vs AO & 12 & 2 & .0129 & .0323 \\
\midrule
\multicolumn{5}{l}{\emph{Direct between-orderings contrast}}\\
AO vs OA & 5 & 13 & .0963 & .1444 \\
\bottomrule
\end{tabular}
\end{table}

\subsection{RQ2: Writer--Reviewer Role Assignment}
Table~\ref{tab:mcnemar} reports the corresponding paired McNemar tests. For \codex{} drafts, both reviewers help after BH correction: OA improves over O by 18.1\,pp ($p_{BH}=.0010$) and OO by 12.9\,pp ($p_{BH}=.0222$). For \claude{} drafts, same-model review is ineffective, while cross-model review by \codex{} is significantly worse than \claude{} solo ($p_{BH}=.0456$). AA versus AO is also significant ($p_{BH}=.0323$), indicating that the choice of reviewer matters even when the writer is already the stronger of the two.

The direct AO vs OA contrast does not survive BH correction ($p_{BH}=.1444$), so we do not claim that the two cross-model orderings are statistically separable as standalone conditions on this sample. 

\subsection{RQ3: Fixes versus Regressions}
\label{sec:qualitative}
\begin{figure}[t]
\centering
\resizebox{\columnwidth}{!}{%
\begin{tikzpicture}[
    x=0.054cm,
    y=0.58cm,
    rowlbl/.style={font=\small, anchor=east},
    grouplbl/.style={font=\scriptsize\itshape, text=gray!55!black, anchor=east},
    legendlbl/.style={font=\footnotesize, anchor=west},
    annot/.style={font=\small, anchor=west},
    seg/.style={draw=white, line width=0.4pt}
]
\node[grouplbl] at (-13.5,3.5) {Claude writer};
\node[grouplbl] at (-13.5,1.5) {Codex writer};
\node[rowlbl] at (-2.5,4) {AA};
\node[rowlbl] at (-2.5,3) {AO};
\node[rowlbl] at (-2.5,2) {OA};
\node[rowlbl] at (-2.5,1) {OO};
\draw[seg, fill=green!55] (0,3.75) rectangle (88.8,4.25);
\draw[seg, fill=blue!60] (88.8,3.75) rectangle (91.4,4.25);
\draw[seg, fill=gray!35] (91.4,3.75) rectangle (97.4,4.25);
\draw[seg, fill=red!60] (97.4,3.75) rectangle (100,4.25);
\node[annot] at (102,4) {\textcolor{blue!55!black}{$+$3} / \textcolor{red!60!black}{$-$3}};
\draw[seg, fill=green!55] (0,2.75) rectangle (80.2,3.25);
\draw[seg, fill=blue!60] (80.2,2.75) rectangle (82.8,3.25);
\draw[seg, fill=gray!35] (82.8,2.75) rectangle (88.8,3.25);
\draw[seg, fill=red!60] (88.8,2.75) rectangle (100,3.25);
\node[annot] at (102,3) {\textcolor{blue!55!black}{$+$3} / \textcolor{red!60!black}{$-$13}};
\draw[dashed, gray!55, thin] (-9,2.5) -- (122,2.5);
\draw[seg, fill=green!55] (0,1.75) rectangle (67.2,2.25);
\draw[seg, fill=blue!60] (67.2,1.75) rectangle (89.7,2.25);
\draw[seg, fill=gray!35] (89.7,1.75) rectangle (95.7,2.25);
\draw[seg, fill=red!60] (95.7,1.75) rectangle (100,2.25);
\node[annot] at (102,2) {\textcolor{blue!55!black}{$+$26} / \textcolor{red!60!black}{$-$5}};
\draw[seg, fill=green!55] (0,0.75) rectangle (66.4,1.25);
\draw[seg, fill=blue!60] (66.4,0.75) rectangle (84.5,1.25);
\draw[seg, fill=gray!35] (84.5,0.75) rectangle (94.8,1.25);
\draw[seg, fill=red!60] (94.8,0.75) rectangle (100,1.25);
\node[annot] at (102,1) {\textcolor{blue!55!black}{$+$21} / \textcolor{red!60!black}{$-$6}};
\draw[fill=green!55, draw=green!55] (0,5.55) rectangle (2.8,5.85);
\node[legendlbl] at (3.8,5.7) {pass kept};
\draw[fill=blue!60, draw=blue!60] (31,5.55) rectangle (33.8,5.85);
\node[legendlbl] at (34.8,5.7) {fix};
\draw[fill=gray!35, draw=gray!55] (47,5.55) rectangle (49.8,5.85);
\node[legendlbl] at (50.8,5.7) {fail kept};
\draw[fill=red!60, draw=red!60] (74,5.55) rectangle (76.8,5.85);
\node[legendlbl] at (77.8,5.7) {regression};
\draw[->, thin] (0,0.35) -- (100,0.35);
\node[font=\small, anchor=north] at (50,0.2) {share of tasks (\%), grouped by writer};
\end{tikzpicture}
}
\caption{Per task outcomes ($n=116$): stacked decomposition with conditions grouped by writer. Absolute fix and regression counts are shown to the right of each bar.}
\label{fig:outcomes}
\end{figure}
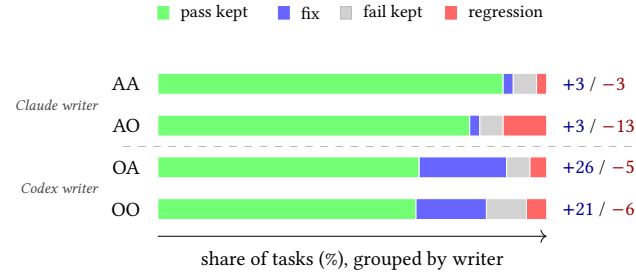

A pass rate hides two different reviewer behaviors: a reviewer can fix a failing draft, or break a passing one. Figure~\ref{fig:outcomes} unpacks them, grouping conditions by writer with explicit fix and regression counts on the right. The contrast that is hardest to miss is between the two cross-model orderings: OA fixes 26 of \codex{}'s failures while regressing only 5 of its successes (net $+21$), and AO does the opposite, fixing only 3 of \claude{}'s failures but breaking 13 of its successes (net $-10$). The two same-model conditions sit between these poles: OO is broadly positive (21 fixes, 6 regressions), and AA is at the noise floor (3 fixes, 3 regressions).

In addition, we read through representative fixes and regressions by hand. The supporting evidence for these examples lives in the artifact records; each record stores the writer artifact, reviewer artifact, final artifact, prompt hashes, and pass or fail outcome. The pattern labels below are interpretive and might not be reproducible in repeated runs.

\begin{table*}[t]
\centering
\caption{Representative inspected cases from artifact records. These examples are not a coded taxonomy but illustrate the mechanisms behind the aggregate fix and regression counts.}
\label{tab:caseexamples}
\setlength{\tabcolsep}{3.5pt}
\renewcommand{\arraystretch}{1.08}
\begin{tabularx}{\textwidth}{@{}lccc>{\raggedright\arraybackslash}X>{\raggedright\arraybackslash}X@{}}
\toprule
Problem & Cond. & Draft & Final & Reviewer action & Interpretation \\
\midrule
3562 & OA & fail & pass & Removed code fences around a dynamic-programming solution so the evaluator could parse the submitted program. & The reviewer corrects a submission failure on parsing. \\
\texttt{abc394\_f} & AO & fail & pass & Replaced an incorrect component-counting solution with a tree dynamic program. & This is a rare helpful AO case. The O reviewer found the missing structural invariant. \\
3688 & OO & fail & pass & Restored deleted values in the segment-tree state and kept the rest of the writer's interface. & The same model O reviewer is helpful when it repairs one local invariant while preserving the draft's architecture. \\
3717 & AO & pass & fail & Replaced a passing sorted-list median-window solution with a heap-based rewrite. & The reviewer discards a correct invariant and introduces a new one that fails hidden tests. \\
\texttt{arc191\_a} & AA & pass & fail & Changed optional digit accounting in an already passing solution. & The same model reviewer can regress when the second pass treats a delicate boundary case as a cleanup opportunity. \\
\texttt{abc389\_f} & OO & pass & fail & Removed a segment-tree range guard that protected an edge case. & A small guard deletion can break a solution that otherwise had the right data structure. \\
\bottomrule
\end{tabularx}
\end{table*}

Table~\ref{tab:caseexamples} shows representative cases at the task level. Fixes include Problem 3688, where OO repairs its own segment-tree state, and \texttt{abc394\_f}, where the O reviewer replaces a failing component count with a tree dynamic program. Regressions include Problem 3717 (AO replaces a passing sorted-list median window with a failing heap-based rewrite), \texttt{arc191\_a} (AA rewrites the boundary handling of an already passing solution), and \texttt{abc389\_f} (OO deletes a segment-tree range guard protecting an edge case).

\subsection{RQ4: Cost and Latency Trade-Offs}
\begin{figure*}[t]
\centering
\resizebox{0.78\textwidth}{!}{%
\begin{tabular}{@{}c@{\hspace{0.9cm}}c@{}}
\begin{tikzpicture}[
    x=19cm,
    y=12.3cm,
    point/.style={circle, draw, fill=white, minimum size=6pt, inner sep=0pt},
    review/.style={diamond, draw, fill=gray!15, minimum size=7pt, inner sep=0pt},
    cA/.style={point, draw=condA, fill=condA!18},
    cO/.style={point, draw=condO, fill=condO!18},
    cAO/.style={review, draw=condAO, fill=condAO!18},
    cOA/.style={review, draw=condOA, fill=condOA!18},
    cAA/.style={review, draw=condAA, fill=condAA!18},
    cOO/.style={review, draw=condOO, fill=condOO!18},
    axis/.style={->, thick},
    grid/.style={gray!25, thin},
    tick/.style={black, thin}
]
\foreach \x/\lab in {0.20/{\$0.20},0.30/{\$0.30},0.40/{\$0.40}} {
    \draw[grid] (\x,0.65) -- (\x,1.0);
    \draw[tick] (\x,0.65) -- (\x,0.635) node[below, font=\scriptsize] {\lab};
}
\foreach \y/\lab in {0.65/{65\%},0.75/{75\%},0.85/{85\%},0.95/{95\%},1.00/{100\%}} {
    \draw[grid] (0.15,\y) -- (0.47,\y);
    \draw[tick] (0.15,\y) -- (0.145,\y) node[left, font=\scriptsize] {\lab};
}
\draw[axis] (0.15,0.65) -- (0.48,0.65);
\draw[axis] (0.15,0.65) -- (0.15,1.03) node[above, font=\footnotesize] {pass rate};
\draw[-{Latex[length=2mm]}, condO!70!black, thick, shorten >=3pt, shorten <=3pt] (0.190,0.716) -- (0.312,0.845);
\draw[-{Latex[length=2mm]}, condO!70!black, thick, shorten >=3pt, shorten <=3pt] (0.190,0.716) -- (0.443,0.897);
\draw[-{Latex[length=2mm]}, condA!70!black, thick, shorten >=3pt, shorten <=3pt] (0.226,0.914) -- (0.382,0.828);
\draw[-{Latex[length=2mm]}, condA!70!black, thick, shorten >=3pt, shorten <=3pt] (0.226,0.914) -- (0.389,0.914);
\node[cA, label={[font=\scriptsize,text=condA]above:A}] at (0.226,0.914) {};
\node[cO, label={[font=\scriptsize,text=condO]below:O}] at (0.190,0.716) {};
\node[cAO, label={[font=\scriptsize,text=condAO]right:AO}] at (0.382,0.828) {};
\node[cOA, label={[font=\scriptsize,text=condOA]right:OA}] at (0.443,0.897) {};
\node[cAA, label={[font=\scriptsize,text=condAA]above:AA}] at (0.389,0.914) {};
\node[cOO, label={[font=\scriptsize,text=condOO]below:OO}] at (0.312,0.845) {};
\node[font=\footnotesize] at (0.315,0.58) {cost/task};
\node[font=\footnotesize] at (0.315,0.53) {(a) Cost};
\end{tikzpicture}
&
\begin{tikzpicture}[
    x=0.058cm,
    y=12.3cm,
    point/.style={circle, draw, fill=white, minimum size=6pt, inner sep=0pt},
    review/.style={diamond, draw, fill=gray!15, minimum size=7pt, inner sep=0pt},
    cA/.style={point, draw=condA, fill=condA!18},
    cO/.style={point, draw=condO, fill=condO!18},
    cAO/.style={review, draw=condAO, fill=condAO!18},
    cOA/.style={review, draw=condOA, fill=condOA!18},
    cAA/.style={review, draw=condAA, fill=condAA!18},
    cOO/.style={review, draw=condOO, fill=condOO!18},
    axis/.style={->, thick},
    grid/.style={gray!25, thin},
    tick/.style={black, thin}
]
\foreach \x/\lab in {40/{40s},80/{80s},120/{120s}} {
    \draw[grid] (\x,0.65) -- (\x,1.0);
    \draw[tick] (\x,0.65) -- (\x,0.635) node[below, font=\scriptsize] {\lab};
}
\foreach \y/\lab in {0.65/{65\%},0.75/{75\%},0.85/{85\%},0.95/{95\%},1.00/{100\%}} {
    \draw[grid] (30,\y) -- (145,\y);
    \draw[tick] (30,\y) -- (28,\y) node[left, font=\scriptsize] {\lab};
}
\draw[axis] (30,0.65) -- (148,0.65);
\draw[axis] (30,0.65) -- (30,1.03) node[above, font=\footnotesize] {pass rate};
\draw[-{Latex[length=2mm]}, condO!70!black, thick, shorten >=3pt, shorten <=3pt] (38.5,0.716) -- (67.9,0.845);
\draw[-{Latex[length=2mm]}, condO!70!black, thick, shorten >=3pt, shorten <=3pt] (38.5,0.716) -- (112.4,0.897);
\draw[-{Latex[length=2mm]}, condA!70!black, thick, shorten >=3pt, shorten <=3pt] (86.2,0.914) -- (118.0,0.828);
\draw[-{Latex[length=2mm]}, condA!70!black, thick, shorten >=3pt, shorten <=3pt] (86.2,0.914) -- (135.8,0.914);
\node[cA, label={[font=\scriptsize,text=condA]above:A}] at (86.2,0.914) {};
\node[cO, label={[font=\scriptsize,text=condO]below:O}] at (38.5,0.716) {};
\node[cAO, label={[font=\scriptsize,text=condAO]below right:AO}] at (118.0,0.828) {};
\node[cOA, label={[font=\scriptsize,text=condOA]right:OA}] at (112.4,0.897) {};
\node[cAA, label={[font=\scriptsize,text=condAA]above:AA}] at (135.8,0.914) {};
\node[cOO, label={[font=\scriptsize,text=condOO]below:OO}] at (67.9,0.845) {};
\node[font=\footnotesize] at (89,0.58) {latency/task};
\node[font=\footnotesize] at (89,0.53) {(b) Latency};
\end{tikzpicture}
\end{tabular}
}
\caption{Cost versus accuracy (left) and latency versus accuracy (right) trade-offs. Circles are solo baselines; diamonds are reviewed conditions; arrows show what each added reviewer buys. \textcolor{condO!70!black}{Red} arrows leave the Codex solo point; \textcolor{condA!70!black}{blue} arrows leave the Claude solo point. Up is more accurate; right is more expensive or slower.}
\label{fig:tradeoff}
\end{figure*}
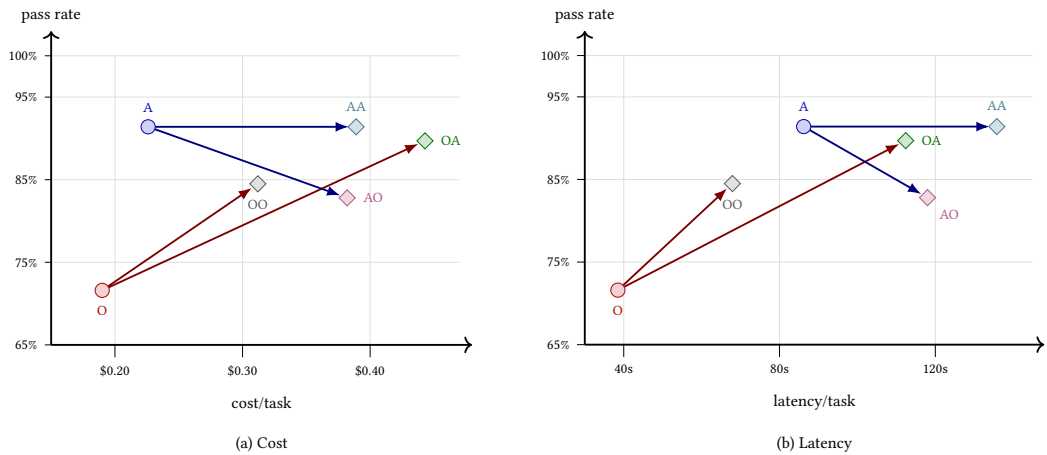

A review pass is not free, as it doubles the API call count and adds tens of seconds of wall-clock latency. The cost-per-token and latency overhead of additional LLM passes has motivated a broad line of work on efficient LLM deployment \cite{cheng2026toward}. Figure~\ref{fig:tradeoff} shows whether that extra spend buys anything. Each arrow starts at a writer's solo point and ends at one of its reviewed conditions, so the slope of the arrow tells you whether the extra cost or latency on the horizontal axis bought accuracy on the vertical axis.

For \codex{} as writer, both arrows go up and to the right. OO buys 12.9 pass rate points for an extra \$0.12 per task and 29 seconds of latency; OA buys 18.1 points for an extra \$0.25 and 74 seconds. OO is the cheap, fast option. For \claude{} as writer, neither arrow justifies the spend. $A\!\to\!AA$ is flat-right: \$0.16 and 50 seconds for zero accuracy gain. $A\!\to\!AO$ points down-and-right: \$0.16 and 32 seconds for an 8.6-point accuracy loss.

Overall, \claude{} solo sits on the Pareto frontier and no reviewed condition matched or exceeded it in accuracy. AA matches A's pass rate while adding 72\% to cost and 58\% to latency; every other reviewed condition has lower accuracy than A. The cost versus accuracy decision worth thinking about is therefore confined to the \codex{} writer column.

\section{Discussion}


\subsection{Claude self review versus Codex self review}
The OO and AA rows in Table~\ref{tab:completecase} and Figure~\ref{fig:outcomes} apply the same operation, ``review your own draft'', to the two writers, and yet produce opposite outcomes. The most plausible explanation is that the two models do different amounts of internal verification on the first pass. At high reasoning effort, \claude{} may already catch most errors that a static reviewer could find, leaving the second pass with little signal. \codex{} appears to leave more residual, catchable errors. 

The latency table supports this reading: \claude{} solo takes 86.2\,s per task against \codex{}'s 38.5\,s, consistent with \claude{} spending more compute on first pass checks. If review value depends on the gap between what the writer already verified and what the reviewer can add, then AA has little to add and OO has a real margin to work with.

\subsection{Claude reviewing Codex helps whereas Codex reviewing Claude hurts}
\label{sec:discuss-cross}
The same asymmetry shows up across the two cross-model conditions in Table~\ref{tab:heatmap}: OA gains while AO loses on the same task set with the same prompts. 

There are two possible hypotheses on this. First, \claude{}'s heavier first pass verification leaves less for any second pass to find, so a \codex{} reviewer staring at a \claude{} draft has few real catches available and tends to fall back on rewriting. Second, \claude{} starts from a higher pass rate baseline, so a \claude{} reviewer has room to lift \codex{}, while a \codex{} reviewer can mostly only drop \claude{}. Both are consistent with the Olausson et al.\ finding that the gain from a repair pass depends on whether the repairer is stronger than the drafter \cite{olausson2024self}: OA is the case where the reviewer is stronger than the drafter, and AO is the reverse. Because the two models differ in baseline pass rate, this design cannot fully separate the effect of review direction from the effect of that baseline gap, so we read the cross-model comparison as descriptive of role assignment rather than as a causal claim about ordering alone.

\subsection{Helpful review edits whereas harmful review rewrites}
\label{sec:discuss-mechanism}
Across the cases in Section~\ref{sec:qualitative}, the OA and AO reviewers differ less in how often they intervene than in how do they intervene. \codex{} as reviewer, when uncertain, tends to discard the writer's data structure and start over (the 3717 heap rewrite and the \texttt{abc389\_f} guard deletion are characteristic). \claude{} as reviewer tends to keep the writer's interface and repair one local invariant, such as 3688 on a segment-tree fix. 

We did not measure rewrite frequency directly, so this is an interpretive pattern across the artifact rather than a quantitative claim. It suggests a concrete prompt direction for future work: instruct the reviewer to first decide whether to intervene at all, and to default to keeping the writer's structure when in doubt.

\subsection{When to add a reviewer, and when not to}
\label{sec:rule}
Cost and latency turn the discussion above into a practical guideline for this benchmark based on the sample data:

\begin{itemize}\setlength{\itemsep}{0.25em}
\item \emph{If the writer must be \codex{}}: use OA when accuracy dominates (89.7\%, $+18.1$\,pp over \codex{} solo, lowest reviewed regression rate at 4.3\%) and use OO when the per task latency budget is under roughly 70\,s and 84.5\% is acceptable.
\item \emph{If the writer must be \claude{}}: skip review. AA matches A's pass rate while adding cost and 49.6\,s of latency; AO loses 8.6\,pp at higher cost than either solo baseline.
\item \emph{Cost per net fix}: OA adds 21 net passes at an extra \$0.25 per task across 116 tasks, or roughly \$1.40 per net fix; OO adds 15 net passes at an extra \$0.12 per task, or roughly \$0.95 per net fix. AO loses 10 net passes at higher cost than either solo baseline, so it has no useful cost per fix number.
\item \emph{If both ends are free}: \claude{} solo is Pareto-optimal on this benchmark. No reviewed condition beats it on both axes, and AA is strictly dominated by A.
\end{itemize}

\subsection{Scope, Limitations, and Reproducibility}
This study is designed for exploratory understanding rather than breadth. At 116 paired tasks drawn from hard and medium \lcb{} code generation problems, the design is sufficient for a first diagnostic but not for settling stable rankings across future model releases. The natural follow-up is to repeat the design on larger \lcb{} dataset as later releases appear. Evaluating the same design on a larger task sample and on additional code benchmarks beyond \lcb{} would also help test whether these patterns generalize. The setting isolates algorithmic reasoning under contamination-aware evaluation using LiveCodeBench but does not generalize to repository-scale bug fixing, build systems, or multi-file review. The static reviewer cannot execute tests, which keeps the setting close to pre-CI code review but likely understates what a tool-using agent with a sandbox could achieve. Results are also sensitive to prompt wording and formatting instructions, and because coding agents change quickly we pin model identifiers, CLI versions, and execution dates, though future releases may shift the numbers in either direction. Cost figures are point-in-time estimates subject to pricing and caching changes.

Our prompts are written manually, which can advantage or disadvantage either model. Stronger or more conservative instructions might reduce harmful rewrites and further tool access could change both the fix rate and the regression rate. There can be further investigation into how prompts can be worded differently and whether models can benefit from memory of past failures, such as prior success and failure review cases, test driven development, or feedback on draft results. A related limitation is structural: in our prompt the reviewer always emits a final program, so there is no separate ``intervene?'' decision. AO's high regression rate (11.2\%) is consistent with \codex{} over-using that rewrite affordance. A prompt that asks the reviewer to first decide whether to intervene, and only then to write a final program, is a natural follow-up.

We used "high" reasoning effort in \codex{} and \claude{}; "extra high" in \codex{} and, more recently, "Max" in \claude{} are now available. Given we were initially attempting to understand whether the review step helps and which model is the better reviewer, we did not focus on different reasoning parameters. In addition, new models are also released very often. It is likely that newer versions of \claude{} and \codex{} have improved results. Further experiments and analysis using the latest models and different reasoning efforts can provide more granular recommendations for software practitioners. Our evidence also covers a single model pair, so whether the asymmetry generalizes beyond \claude{} and \codex{} to families such as Gemini, DeepSeek, Qwen, or Grok is untested, and repeating the design across more models, including within one family, is a direct way to check for consistent patterns. 

The benchmark tasks are self contained Python programs with hidden tests, not patches inside a live repository. This makes correctness measurement precise, but it does not consider areas such as API compatibility, build integration, style consistency, security posture, and maintainability across files. Restricting to self contained functions is what isolates the writer and reviewer roles from repository context and multi file dependencies, which is why we treat it as a first controlled diagnostic before repository scale review. We therefore view the results as evidence about model role assignment under controlled static review, not as a universal ranking of either model as a software reviewer. We also score review only by the correctness of the final program. Code review in practice yields more than a pass or fail signal, including design feedback, readability comments, and security observations, and whether LLM reviewers add value along those dimensions is left to future work.

Lastly, for reproducibility, the released artifact\footnote{Code and data: https://github.com/shawnzxiang/cross-model-review-code} includes the full experiment configuration, prompts, raw outputs, prompt hashes, CLI versions, timing and token fields, and analysis scripts. Reproducing all six writer-reviewer conditions across the 116 complete case tasks costs roughly \$225 in API calls (or \$1.94 per task across 6 conditions, see Table~\ref{tab:completecase}), with retries, models, and reasoning effort configurable.

\section{Conclusion}
We started with a simple question: if you have two coding agents and let one write a draft while the other reviews it before the code is submitted, does the review pass help? Using \claude{} and \codex{}, we ran every writer and reviewer combination on 116 hard and medium \lcb{} programming problems. The reviewer saw the problem and the draft but could not run the code or see test results. 

The answer depends on who wrote the draft. When \codex{} writes, a review pass helps: \claude{} reviewing \codex{} raises the pass rate from 71.6\% to 89.7\% (a gain of 18.1 points), and \codex{} reviewing its own draft raises it to 84.5\% (a gain of 12.9 points). When \claude{} writes, a review pass does not help. \claude{} reviewing itself leaves its 91.4\% pass rate unchanged, and \codex{} reviewing \claude{} drops it to 82.8\% (a loss of 8.6 points). Looking at individual tasks shows that a helpful review fixes a small flaw in an otherwise sound draft, while a harmful review throws away a working draft and rewrites it into one that fails. 

The implication in this paper is practical for software development. If \codex{} writes the draft, add a review pass, and use \claude{} as the reviewer for the best accuracy or \codex{} itself when cost and latency matter more. If \claude{} writes the draft, submit it as is, because no reviewer we tested beats \claude{} working alone. The exact numbers are specific to these two model versions and to \lcb{}. Future research on areas such as modifying reviewer prompts, scaling up the task sample, evaluating on additional code benchmarks beyond \lcb{}, and testing a wider range of LLM models and configurations can provide us with a deeper understanding of review agents in the era of Agentic AI software development. 

\bibliographystyle{ACM-Reference-Format}
\bibliography{references}

\end{document}